\documentclass{aa}
\usepackage{graphicx}
\usepackage{txfonts}
\usepackage[]{natbib}
\newcommand{\spitzer}{{\it Spitzer}}
\newcommand{\mi}{{$\mu$m}}

\begin{document}
   \title{The cluster birthline in M33}

   \author{Edvige Corbelli
          \inst{1}
          \and
          Simon Verley
          \inst{1}
          \and
          Bruce G. Elmegreen
          \inst{2}
          \and
          Carlo Giovanardi
          \inst{1}
          }

%   \offprints{}

   \institute{INAF-Osservatorio Astrofisico di Arcetri, Largo E. Fermi, 5 - 50125
              Firenze - Italy\\
              \email{[edvige, simon, giova]@arcetri.astro.it}
         \and
              IBM Research Division, T.J. Watson Research Center, 1101 Kitchawan Road, 
              Yorktown Hts., NY 10598\\
              \email{bge@us.ibm.com}
             }

   \date{Received; accepted}

% \abstract{}{}{}{}{}
% 5 {} token are mandatory

  \abstract
   {}
   {The aim of this paper is twofold: $(a)$ to determine the reliability of
   infrared (IR) emission to trace star formation in individual star-forming sites 
   of M33, and $(b)$ to outline a new method for testing the distribution function 
   of massive stars in newly formed clusters.
   }
   {We select 24~\mi\ IR sources from the
   \spitzer\ survey of M33 with H$\alpha$ counterparts and show that the IR 
   luminosities have a weak dependence on galactocentric radius.  
   The IR and H$\alpha$ luminosities are not correlated.
   Complementing the infrared
   photometry with GALEX-UV data, we estimate the bolometric luminosities
   to investigate how they are related to the H$\alpha$ luminosities.
   We simulate a theoretical diagram for the expected bolometric-to-H$\alpha$
   luminosity ratio, L$_{bol}$/L$_{H\alpha}$, of young clusters as a function of the
   cluster luminosity. We then compare the observed L$_{bol}$/L$_{H\alpha}$ ratios
   with the theoretical predictions.
   }
   {In the log(L$_{bol}$)--log(L$_{bol}$/L$_{H\alpha}$) plane, stellar clusters
   should be born along a curve that we call the {\it cluster birthline}.
   The birthline depends on
   the stellar initial mass function (IMF) at the high-mass end, but not on 
   the cluster mass function.
   For an upper stellar mass limit of $120\;M_\odot$,
   the birthline is flat for L$_{bol}>3\times10^{39}$~erg~s$^{-1}$ because all 
   clusters fully sample the IMF. It increases toward lower luminosities as the 
   upper end of the IMF becomes incompletely sampled.
   Aging moves clusters above the birthline.
   The observations of M33 show that young isolated clusters lie close to the 
   theoretical birthline for a wide range of L$_{bol}$.
   The observed L$_{bol}$/L$_{H\alpha}$ ratio increases toward low L$_{bol}$ 
   like the theoretical curve, indicating that luminosity
   is not proportional to H$\alpha$ emission for low mass clusters.
   The best fit to the
   birthline is for a randomly sampled IMF, in which the mass of most massive 
   star in a cluster
   is not strictly limited by the cluster's mass, but can have any value up 
   to the maximum stellar mass with a probability determined by the IMF.
   We also find that the IR luminosity of young stellar clusters in M33 is not
   proportional to their bolometric luminosity. This irregularity could be the 
   result of low and patchy dust abundance. In M33 dust 
   absorbs and re-radiates in the IR only part of the UV light from young clusters.
   }
  % conclusions heading (optional)
  {}


   \maketitle
%
%%%%%%%%%%%%%%%%%%%%%%%%%%%%%%%%%%%%%%%%%%%%%%%%%%%%%%%%%%%%%%%%%%%%%%%%%%%%%%%

\section{Introduction}

%%%%%%%%%%%%%%%%%%%%%%%%%%%%%%%%%%%%%%%%%%%%%%%%%%%%%%%%%%%%%%%%%%%%%%%%%%%%%%%

Our knowledge of interstellar conditions that favor the
birth of stars and clusters is mostly based on the Milky Way, but
these studies are ill-suited for examining
massive stars, which are an embedded population with a short lifetime.
The number of Milky Way
clusters that one can sample is also limited by extinction
and distance uncertainties.  These problems are somewhat diminished in 
studies of other galaxies, where imaging with high sensitivity
and resolution is possible thanks to
space observatories such as Spitzer, HST and GALEX. Soon we shall be able
to know if the IMF and extinction in star-forming sites varies
along the Hubble
sequence, and if ISM perturbations from
these sites vary with cluster position and mass.
In the mean time, the best galaxies for study are in the Local Group.
M33, at a distance $D=840$~kpc \citep{1991ApJ...372..455F}, has a higher SFR
per unit area than M31 \citep[3.4 versus 0.74~M$_{\odot}$~Gyr$^{-1}$~pc$^{-2}$,]
[]{1998ApJ...498..541K} and a lower extinction towards star-forming regions
because of its lower inclination
\citep[e.g.][and references therein]{1980ApL....21....1I,2007A&A...470..865M}.
Bearing no prominent bulge and no signs of recent mergers, M33 is a prototype for
star formation in a quiescent galaxy and for evolutionary
scenarios in blue, low luminosity systems. Observations of the molecular,
atomic and ionized gas in M33 have been carried out in the past years
\citep[e.g.][]{1987A&AS...67..509D,1997ApJ...479..244C,2000ApJ...541..597H,
2003MNRAS.342..199C,2003ApJS..149..343E,2004ApJ...602..723H}.
Together with recent works on the metallicity content of the young and
old stellar population across the M33 disk, they allow us to trace the global 
formation and evolution history of the nearest blue disk galaxy
\citep[][ and references therein]{2007A&A...470..843M}.

\citet{2007A&A...476.1161V,2008arXiv0810.0473V}
have shown that the recent {\it Spitzer} images of M33 allow us
to study the average radial variation of star formation properties across the disk,
and to locate stellar clusters and individual stars with
infrared luminosities as low as $2\times 10^{37}$~erg~s$^{-1}$ (corresponding to
a B2 spectral type star on the main sequence). M33 therefore gives us
a unique opportunity to study the properties of many individual star-forming
sites, containing a range of populations from
single O and B-type stars to young stellar clusters and associations. The observable 
luminosity range
extends up to $(4.5\pm 1.5)\times 10^7\;L_\odot$ for NGC 604 
\citep{1982A&A...105..229I,1986A&AS...66..117I,2003A&A...407..137H}, 
the second brightest HII complex in the Local Group.

The aim of this paper is to examine the conditions of the interstellar medium
where stellar clusters are born, to define stellar
cluster properties related to the birth and death of massive stars,
and to examine star formation rate (SFR) diagnostics of individual complexes in M33.
We will show the importance of complementing infrared photometry with photometry at
optical and ultraviolet wavelengths. Previous papers outlined the
need for infrared photometry to supplement H$\alpha$ in determining the SFR
\citep[e.g.][ for M51]{2005ApJ...633..871C,2007ApJ...671..333K}.
In bright star-forming regions of more distant and brighter galaxies,
such as M51, there is a tight linear relation between IR luminosity and the 
strength of extinction-corrected, optical recombination lines
\citep[e.g.][]{2005ApJ...633..871C}. On the other hand, in late-type galaxies 
there are strong variations in the H$\alpha$-to-infrared flux ratio, which limits 
our ability to trace star formation via infrared
photometry alone \citep[e.g.][ for the metal deficient Local dwarf
NGC6822, where the H$\alpha$-to-IR flux ratio
variations are as high as a factor 10]{2006ApJ...652.1170C}.
UV observations are also needed
to trace the properties of star formation, particularly where the HII regions are 
density bounded or the O-type stars have evolved off the main sequence.
GALEX observations of M33 have shown that the UV
luminosity of a region is comparable to the IR luminosity \citep{2008arXiv0810.0473V},
and is therefore a significant contribution to the total.

Here we employ a new
technique to extract and measure sources from M33 images.
Because dust and its associated IR emission are distributed in both star-forming 
complexes and the diffuse interstellar medium,
we do photometry with varying apertures to match the source size at each wavelength.
We also combine IR, UV and H$\alpha$ fluxes to give source properties
down to very faint emission levels.

It is currently unknown if IR fluxes are good diagnostics for
the SFR in areas where the bolometric luminosity of the newborn cluster
is below 10$^{40}$~erg~s$^{-1}$. M33 is an excellent candidate for studying 
individual low-mass clusters because its proximity minimizes
contamination from the surrounding ISM.
Whether low-mass clusters contribute substantially to the SFR of a galaxy
depends on the initial cluster mass function (ICMF).
In M33 the fragmentation process seems to favor
small masses: the mass spectrum slope for giant molecular clouds (GMCs) is steeper
than $-2$, as is
the HII region luminosity function determined from H$\alpha$ emission
\citep[e.g.][]{1997PASP..109..927W}. \citet{2007A&A...476.1161V} have shown that
the luminosity function at 24~$\mu$m  can be described by a double power-law
that is shallower than $-2$ at its faint end. While a change of slope like this 
could mark the transition between poor and rich clusters, occurring at the
luminosity of the brightest star formed, a shallow slope at the low
luminosity end can also indicate a decreasing dust abundance for lower masses. 
Giant complexes may form primarily in dust-rich environments, e.g. close to spiral 
arms, while low-mass clusters may form everywhere, also in low-dust environments.
If some low-mass clusters are born in a low or patchy dust environment, then we cannot
use IR luminosities alone to infer the SFR.

In this paper we will study the large
scatter in the IR-to-H$\alpha$ ratio that is observed  when young, low-luminosity
stellar clusters are selected in M33. We show that
complementing the IR photometry with UV photometry helps recover the total
cluster luminosity and  improves the correlation between
stellar continuum and gas recombination lines from star-forming regions.
We will examine in detail the theoretical and observed relation
between L$_{bol}$ and L$_{H\alpha}$, the deviations from the linear
regime, and some  properties of the massive stellar population in young clusters.

We shall also investigate the 8\mi\ emission from PAHs in individual HII regions.
Generally, PAH emission is low in dwarf and low-luminosity blue galaxies 
compared to other IR radiation. This may be related
to the low metallicity and intense stellar radiation field in these
systems relative to those in spiral galaxies
\citep[e.g. ][]{2003A&A...407..159G,2003ApJ...588..199L,2004ApJS..154..211H,
2005ApJ...624..162H,2006ApJ...636..742R}. \citet{2001ApJ...553..121H}
for example, used Infrared Space Observatory (ISO) mid-IR imaging and
far-IR (FIR) spectroscopy to examine the properties of five dwarf
irregular systems. They found that PAH emission, which is associated only
with the brightest H II regions, is depressed relative to that of small
grains and far-IR. In addition, the integrated [C II] emission
relative to PAH emission is high in dwarfs compared to spiral
galaxies, suggesting that atomic carbon is elevated
relative to PAHs in dwarfs. \citet{2005ApJ...628L..29E} also examined
low-metallicity systems and found that 8.0\mi\
emission decreases abruptly relative to 24\mi\ dust emission when
the metallicity is less than one-third to one-fifth solar.
M33 is a low-luminosity galaxy hosting HII regions with metallicities
between solar and one-fifth solar and a shallow radial
gradient \citep{2007A&A...470..865M,2008ApJ...675.1213R}.
PAHs are not globally underabundant
in M33 although there is a faster decline of 8\mi\ emission
relative to longer wavelengths beyond 3.5~kpc
\citep{2008arXiv0810.0473V}.
It is of interest then to examine possible metallicity and
radial dependences of PAH features in individual HII regions.

In Section 2 we define young stellar cluster samples in M33
and analyze their IR properties. In
Section 3 we study their UV luminosities and colors and define the
cluster bolometric luminosities. The concept of {\it cluster birthline} is
given in Section 4 and tested using M33 young clusters.
In Section 4 we discuss the implications of our results relative to 
the dust abundance, to the bolometric to H$\alpha$ luminosity ratio,
and to the population of massive stars in 
star-forming sites of M33. In the Appendix we analyze
possible metallicity dependencies using a small sample of young clusters
for which gas metal abundances are known, as well as dependencies of 
cluster properties on the mass of the associated giant molecular cloud.

%%%%%%%%%%%%%%%%%%%%%%%%%%%%%%%%%%%%%%%%%%%%%%%%%%%%%%%%%%%%%%%%%%%%%%

\section{Multiwavelength observations of star-forming sites}

%%%%%%%%%%%%%%%%%%%%%%%%%%%%%%%%%%%%%%%%%%%%%%%%%%%%%%%%%%%%%%%%%%%%%%%

The closest star-forming galaxies offer a unique opportunity
to resolve individual star-forming regions. In M33,
the star formation process can be investigated not only by
averaging over selected areas of the disk or by using very bright complexes,
but also by observing individual star clusters of moderate size. The sensitivity 
of telescopes such as Spitzer in the IR and GALEX in the UV allows us to
sample clusters over 4 orders of magnitude in brightness: from the second
brightest HII region of the Local Group (NGC 604) down to clusters of about 1000
L$_\odot$. We can look in detail at the
site properties  where clusters of different masses are born,
and at the emission properties of young clusters.
We will use photometric information on sources
across the M33 disk as given by the Spitzer 8$\mu$m and 24$\mu$m images
described in \citet{2007A&A...476.1161V}, by the GALEX FUV and NUV images
from the GALEX Atlas \citep{2007ApJS..173..185G}, and by the H$\alpha$ image
of \citet{2000ApJ...541..597H}.

\subsection{The main sample}

We define a sample of young stellar clusters and isolated massive stars
using the IR catalog of 515 sources extracted from the 24$\mu$m
Spitzer map by \citet{2007A&A...476.1161V}.
This catalog has a completeness limit of $3\times 10^{36}$~erg~s$^{-1}$
(this limiting luminosity refers to the 90$\%$ level of completeness 
computed by extrapolation of the observed cumulative luminosity function at 
L$_{24} > 10^{37}$~erg~s$^{-1}$).
We remind the reader that the photometry for these sources has
been done with a varying aperture to match the source size.
From the catalog we select only sources
that have an H$\alpha$ counterpart. We search for H$\alpha$ counterparts to
24$\mu$m sources as follows.  We first define a sample of H$\alpha$ sources
by extracting 413 sources from the H$\alpha$ image using the SExtractor
software with no specified sky coordinate list and with varying aperture
\citep[][ specific inputs given in Verley et al. 2007]{1996A&AS..117..393B}.
For any source in the IR catalog we then
define a searching radius, $R_s$, equal to the radius of the source
in the 24$\mu$m image (minor axis if the source PSF is elliptical). This radius
is derived by \citet{2007A&A...476.1161V} using SExtractor in the framework of
the ``hybrid'' method and varies between 12 and 200~pc.
If at least one of the sources in
the H$\alpha$ sample is at a distance less than or equal to the searching radius
of an IR source, then this infrared source is selected and its H$\alpha$
flux is set equal to the sum of the fluxes of all H$\alpha$ sources within
$R_s$. If no H$\alpha$ sources are found within $R_s$, then we attempt to measure
the H$\alpha$ flux using the ``hybrid'' photometric method devised by
\citet{2007A&A...476.1161V}. With the ``hybrid'' photometric method, we search for
H$\alpha$ emission in the circular area of radius $R_s$
around the positions of IR sources (specifying the
source coordinates to the SExtractor software).
If again an H$\alpha$ feature is found within the
searching area, then this source is selected and its H$\alpha$
flux set equal to the ``hybrid'' photometric flux. If H$\alpha$ emission
is found at a distance larger than $R_s$, then the IR source is not selected
for our sample.
The use of the hybrid method allows us to recover many faint H$\alpha$ 
counterparts because their approximate positions are entered into SExtractor.
From both methods we have a final sample of 355 sources that have been selected 
from the 24$\mu$m Spitzer map and have an H$\alpha$ counterpart.
Of these, only 3 have a double counterpart in H$\alpha$, less than 1$\%$.
We estimate the completeness limit of the  survey to be
$3\times 10^{35}$~erg~s$^{-1}$ in H$\alpha$  (90$\%$ level of completeness).

Using the 8$\mu$m Spitzer map, we also search for
8$\mu$m counterparts to all 24$\mu$m sources. We employ the same method used
for finding H$\alpha$ counterparts.
We find that 491 of the 515 sources in the 24$\mu$m catalog
have an 8$\mu$m counterpart. For the subsample of interest in this paper
(the 355 IR sources with an H$\alpha$ counterpart), 349, or almost all, have
a clear detection at 8$\mu$m. To the limit of our completeness for 8$\mu$m,
estimated to be $10^{37}$~erg~s$^{-1}$  (90$\%$ level of completeness), 
we find that 8$\%$ of the sources have multiple counterparts at 8$\mu$m.
If the associated PAHs are located along the photodissociation regions and
HII shells, then more than one 8$\mu$m bright rim might be detected per source
even if one stellar cluster is powering the 24$\mu$m emission at the
center.

In Figure 1 we show that the source sizes
are always much smaller than their average separations, so the sources are distinct.
The scatter around the average source size is small, comparable to the size of the
symbols used in the plot. Source sizes decrease radially outward and
their average separations increase.
The 8\mi\ and H$\alpha$ images are used at
the original resolution, higher than that of the 24\mi\ image.
The use of images smoothed to the spatial resolution of the 24\mi\ image, 6~arcsecs,
would have more firmly constrained the total flux at 8\mi\ and H$\alpha$ for some
cases, but it would also have diluted the fainter sources in the midst of high 
diffuse emission, making them harder to detect.
The presence of elliptical source PSFs also makes it difficult to match the
source fluxes at all wavelengths. Also, for the main sample, the 8\mi\ and H$\alpha$
fluxes might be slightly higher in crowded fields than what we derive
(i.e., we might be missing some multiple faint components with luminosities on the order
of our completeness limit).
However, as we shall see later, this will not affect the main results derived
using the isolated source sample. This sample, defined more precisely later in 
this Section, consists of round, compact sources without multiple counterparts; 
they are found by visually inspecting the images at all wavelengths.

In Figure 2 we show source luminosities, the 24$\mu$m-to-H$\alpha$ ratio and the
24$\mu$m-to-8$\mu$m ratio as functions of  the galactocentric
radius $R$. The 24$\mu$m and 8$\mu$m luminosities
are defined as $\nu L_\nu$, where $L_\nu$ is the luminosity per unit frequency at
24 and 8$\mu$m, i.e. $L_\nu=4\pi D^2 F_\nu$, $F_\nu$ being the IR flux measured
and $D$ the distance to M33.
$L_{24}$, $L_{8}$, and $L_{H\alpha}$ (uncorrected for
extinction) are in units of erg s$^{-1}$.
The ratio $L_{24}/L_{H\alpha}$ has a high dispersion and is consistent with
having no radial dependence. The slope of the linear fit in the
log($L_{24}/L_{H\alpha})-R$ plane
is $-0.06\pm 0.02$ with a Pearson linear correlation coefficient
of -0.14. Brighter IR sources lie preferentially at smaller radii,
but the radial dependence is not strong enough to be dominant when the ratio of 24\mi\
to H$\alpha$ luminosity is considered. H$\alpha$ emission appears to
have a very marginal radial dependence and a higher dispersion than the IR luminosities.
If extinction corrections, generally higher at smaller radii, were applied to the 
H$\alpha$ source luminosities, then the sign of the poor correlation found between 
$L_{24}/L_{H\alpha}$ and $R$ might be reversed.
However, we show in the rest of this paper that extinction corrections are
small; M33 is a galaxy with low dust content \citep[see also ][]{2008arXiv0810.0473V}.

\begin{figure}
\includegraphics[width=\columnwidth]{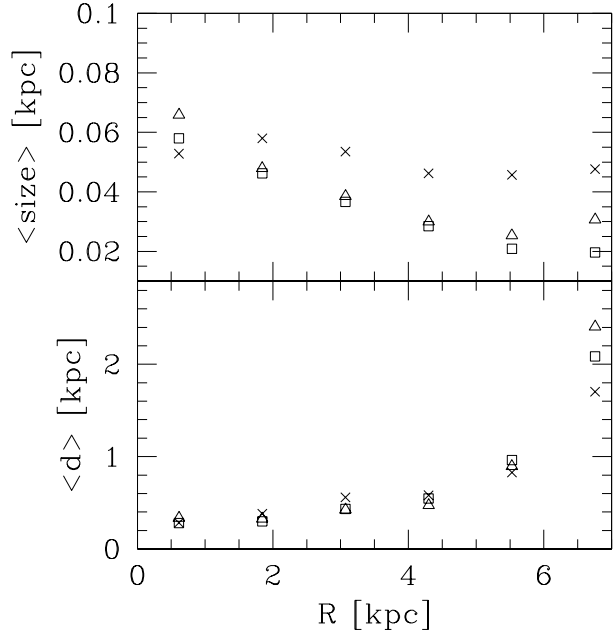}
\caption{The average distance between sources, $d$, and the average source size, $s$,
are shown as a function of galctocentric radius. The dispersions around the average
source sizes are comparable to the symbol sizes. Different symbols refer to measurements
at different wavelengths: 24\mi\ (open triangles), 8\mi\ (open squares), H$\alpha$
(crosses).}
\label{fig:figure1}
\end{figure}

\begin{figure}
\includegraphics[width=\columnwidth]{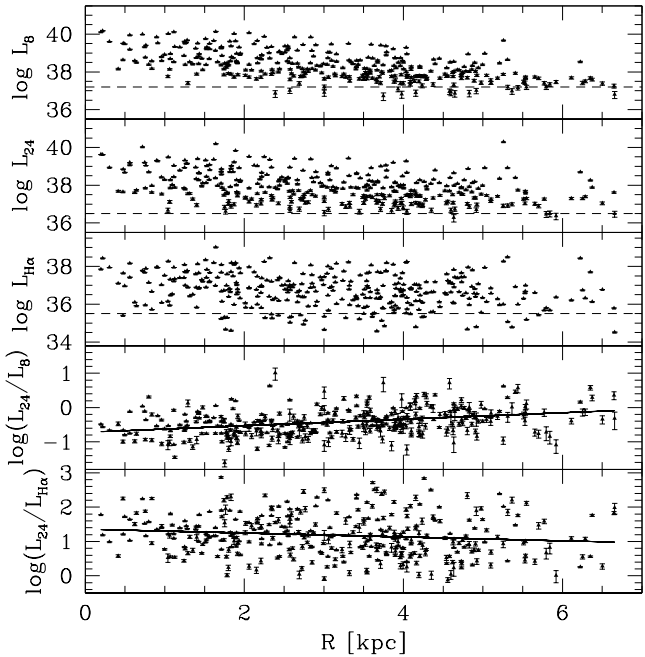}
\caption{The ratio $L_{24}/L_{H\alpha}$, $L_{24}/L_{8}$ and the
H$\alpha$, 24\mi\ and 8\mi\ luminosities (in units of erg~s$^{-1}$)
for IR selected sources in M33 are shown as functions of their distances from the
galaxy center. The dashed lines in the upper 3 panels show the survey completeness
limits. The continuous lines in the bottom 2 panels are the best linear fits
to the data.}
\label{figure2}
\end{figure}

The ratio $L_{24}/L_{8}$ has a  radial dependence that is weak but
undoubtedly exists: the slope of the linear fit shown in Figure 2
is $0.10\pm 0.01$ and the Pearson linear correlation coefficient is
0.34. The radial dependence of the 8$\mu$m source luminosity
is stronger than that of the 24$\mu$m luminosity, so the 24-to-8$\mu$m
luminosity ratio increases at large galactocentric radii.
Because PAHs are likely responsible for the 8$\mu$m emission in M33,
it might be that in the outer disk the lower ISM pressure pushes the
photodissociation regions, where PAHs reside, further from the bulk of the 24$\mu$m
emission. Then our algorithm for finding 8\mi\ counterparts to 24\mi\ sources
would fail at large galactocentric radii.
However, \citet{2008arXiv0810.0473V}
have found that the average radial profile at 8\mi\ in
M33 falls off more steeply than that at 24, 70 and 160\mi, and at UV wavelengths.
Hence PAH carriers seem effectively underabundant at large radii.
There have been claims in other
galaxies that this might be due to the low metal content of the outermost regions,
but it is unclear if this applies to M33 because of its
very shallow metallicity gradient \citep{2007A&A...470..865M}.

\subsection{Total IR luminosities}

\citet{2007A&A...476.1161V} have already pointed
out the large scatter between the SFR
in selected  HII regions of M33
computed via the H$\alpha$ optical recombination
line and that inferred from the 24\mi\ infrared flux.
We shall show that this cannot be
from extinction affecting the H$\alpha$ line because M33 has
a globally low dust content. Nor does it result from
total IR luminosities of star-forming sites in M33 that are not well
represented by 24\mi\ luminosities.

\begin{figure}
\includegraphics[width=\columnwidth]{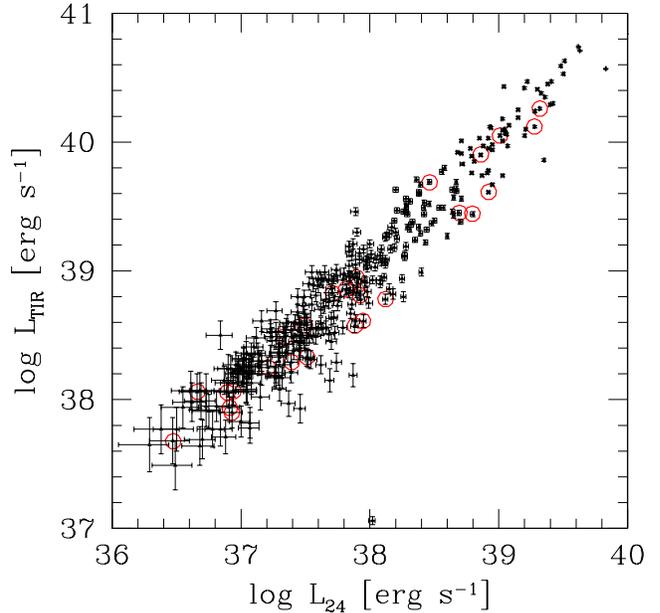}
\caption{The total infrared luminosity of sources in the main sample
as a function of the 24\mi\ luminosity
derived according to Eq.\ref{eq:tir}. Circles mark sources which belong to the
isolated sample. }
\label{fig:tir-24}
\end{figure}

\begin{figure}
\includegraphics[width=\columnwidth]{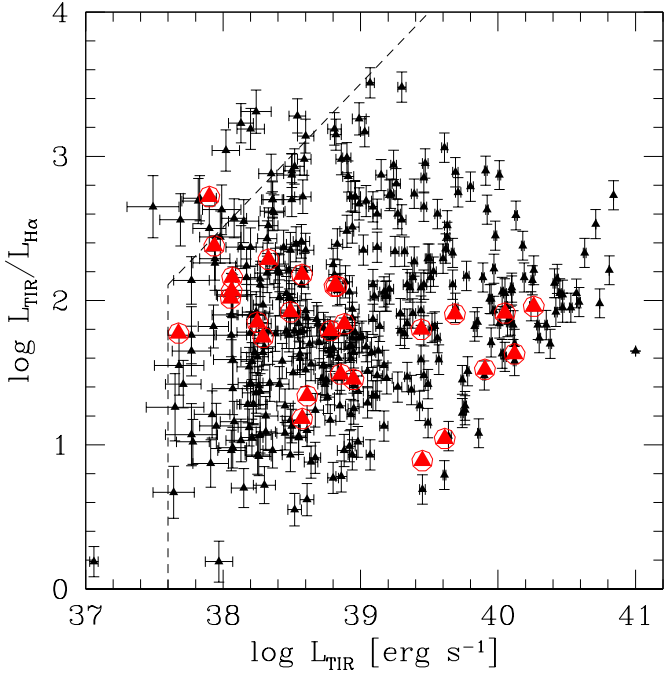}
\caption{Ratio of the TIR
to H$\alpha$ luminosity as a function of the TIR luminosity for IR selected sources.
An average value A$_V=1\pm0.25$~mag has been used to correct the H$\alpha$ luminosity
for extinction. The TIR luminosity has been inferred using Eq.\ref{eq:tir}.
Circles mark sources which belong to the isolated sample. Dashed lines mark
the completeness limits of our survey.
}
\label{fig:tir-ha}
\end{figure}

\citet{2008arXiv0810.0473V} have shown that the total
infrared luminosity (hereafter TIR, defined as the emission between 3-1100 $\mu$m)
in star-forming regions correlates with
the 24\mi\ emission, with a weak additional dependence on the 8-to-24\mi\ ratio.
They show that for a sample of sources selected at 160\mi, the
$\log\left(F_{24}/F_{TIR}\right)$ varies between -0.8 and -1.2 when
$\log\left(F_{8}/F_{24}\right)$ varies between -0.5 and 0.3.
The best fitting relation is

\begin{equation} \label{eq:tir}
\log L_{TIR} = \log L_{24} + 1.08 + 0.51\ \log\left(\frac{F_\nu (8)}{F_\nu
(24)}\right)
\end{equation}

\noindent
where the 24$\mu$m luminosity is $\nu L_\nu$, as usual.
We shall therefore use the above equation to compute the TIR luminosities
of star-forming sites in our main sample. These are shown in Fig.~\ref{fig:tir-24}.
For the 6 sources with no
clear 8\mi\ detection, we neglect the last term in Eq.~\ref{eq:tir}.
In Figure~\ref{fig:tir-ha}, we show the ratio of the TIR
to H$\alpha$ flux as a function of the TIR luminosity for all 355 sources in
the main sample. We encircle sources in the isolated sample (defined in the
next subsection).
The sources span over two orders of magnitude in the TIR to
H$\alpha$ luminosity ratio and about 2.5 orders of magnitude in L$_{TIR}$.
In Figure~\ref{fig:tir-ha} the H$\alpha$
flux has been corrected for extinction using an average A$_V$
of 1~mag (0.83~mag at H$\alpha$).
Since the maximum extinction we measure in the next Section
corresponds to A$_V=1.7$~mag, which is also the average extinction
measured towards bright HII regions \citep{1980ApL....21....1I},
more heavily extincted than faint ones, we consider a dispersion
of 0.25~mag in A$_V$ in the error propagation analysis for
Figure~\ref{fig:tir-ha}. Extinction uncertainties
dominate over photometric errors for bright sources.

The TIR-to-H$\alpha$ luminosity ratio for our main sample
has a large dispersion and shows no dependence on galactocentric radius or
source brightness.
In the next Sextion we will use UV photometry to compute H$\alpha$
extinction corrections for each source in the round and isolated samples. The results
prove that the large scatter we observe in Figure~\ref{fig:tir-ha} is intrinsic
to the sources and is not the result of measurement errors.

We shall show in the rest of the paper that the scatter in Figure~\ref{fig:tir-ha}
is related to three effects:
$(a)$ a varying dust opacity around sources, which implies that the TIR is not
proportional to the cluster bolometric luminosity; $(b)$ a non linear relation
between the bolometric and H$\alpha$ luminosity of newly born clusters
when the cluster luminosity is below a certain threshold, and $(c)$ cluster aging.

\subsection{UV photometry: the round sample and the isolated sample }

We now complement the infrared and H$\alpha$ photometry with the far- and
near-UV data. UV colors can also be used to infer the age of 24$\mu$m
selected sources. GALEX-UV images of M33 are
available \citep{2007ApJS..173..185G} and we shall use them at the original
resolution (1.5~arcsec) to measure the FUV
and NUV luminosities of 24~$\mu$m sources with H$\alpha$ counterparts.
Because we perform
the UV photometry using a circular aperture, we select 106 sources at 24~$\mu$m
using the requirement that the PSF ellipticity is less than 0.2.
We shall call this the {\it round sample}. We set the
radius of the UV photometric aperture equal to the source size at 24~$\mu$m
and consider an annulus with a width equal to 2 pixels (3 arcsec for GALEX) around
sources for background removal.
It is not desirable to perform individual UV source extraction
because of crowding due to the high density of UV sources.
As pointed out already in
previous papers \citep[e.g.][]{2005ApJ...633..871C}, the UV emission peaks are
sometimes displaced with respect to the infrared emission peaks.
Stellar complexes can be made of 2 or more populations of different ages,
both visible in the UV, with the older ones more extended than the
young ones. In this case the old ones should be included in the background.
We would like to
measure only the UV flux coming from the same area covered by IR emission.
To strengthen our results we have selected a clean sample, made of 26 sources
that have a clean annulus in both the FUV and NUV images, i.e. no source
contamination for background subtraction. We shall call this
{\it the isolated sample}.
Briefly, the sources in the isolated sample are sources from the main
sample which satisfy the following criteria:
$(i)$ the PSF of the source at 24$\mu$m is
nearly circular i.e. the ellipticity is $<~0.2$ 
and all of the UV flux is inside $R_s$; $(ii)$ the source
has one H$\alpha$ counterpart (as defined earlier in this paper);
$(iii)$ in the far- and near-UV maps, an annulus of 2 pixels around the
emitting source is devoid of any sources.

We will use the round sample, which includes the isolated sample,
in the rest of the paper to understand some basic properties of young stellar
clusters in M33. 
To check that we are not oversubtracting the background in the case
of sources which do not have a clean annulus, we compute the average UV
luminosities for the isolated sample and for the ensemble of sources
in the round sample which are not in the isolated sample. The ratio between the
two average luminosities is close to unity, being 38.556 and 38.365 
the FUV and NUV logarithmically averaged luminosities (in erg~s$^{-1}$)
in the round and non-isolated
sample, and 38.563 and 38.377 respectively for the isolated sample.
We are  therefore confident in the ability of our technique for UV
background removal to recover the UV luminosities associated with young clusters.

The UV AB-magnitudes were converted to luminosities
using:

\begin{equation}
L_{UV}=\nu F_\nu 4\pi D^2
\end{equation}

\noindent
with $D$ the distance to M33,
$\nu = 1.95 10^{15}$~Hz for the FUV band, $\nu = 1.3 10^{15}$~Hz
for the NUV band, and

\begin{equation}
{F_\nu \over Jy} = 10^{23-(AB+48.6)/2.5}.
\end{equation}

Considering that the GALEX flux calibration has been revised after the early data
release, we apply the recommended 0.12 magnitude offset to the FUV-NUV color
defined as $\log\left[L_\lambda(FUV)/L_\lambda(NUV)\right]$ \citep{2006AJ....132..378B}.
We consider also the small extinction correction, E(B-V)$_{MW}\simeq 0.04$, 
from the Milky Way in the direction of M33 \citep{1998ApJ...500..525S}. This
implies $A_{UV}\simeq 0.3$ for all sources in M33.

%%%%%%%%%%%%%%%%%%%%%%%%%%%%%%%%%%%%%%%%%%%%%%%%%%%%%%%%%%%%%%%%%%%%%%

\section{UV colors, opacity and bolometric luminosities of star-forming regions}

%%%%%%%%%%%%%%%%%%%%%%%%%%%%%%%%%%%%%%%%%%%%%%%%%%%%%%%%%%%%%%%%%%%%%%%

The TIR luminosity of HII regions is radiation from young stars
absorbed and re-emitted by dust grains. The leftover cluster radiation,
not absorbed by grains, is mostly emitted in the UV.
We now infer the extinction of the radiation coming from
the gas and the stars in the HII regions
using the observed infrared-to-far UV ratio.
As in \citet{2001PASP..113.1449C}, we shall use the factor 1.68 to
account for the bolometric correction of the stellar emission relative
to the FUV.

Since our IR and UV data refer to clusters still embedded in HII
regions, the dust optical depth for the stellar continuum radiation is
similar to that for the H$\alpha$ line emission. We are not considering
the UV light from previous generations of stars.
We infer the FUV extinctions after
measuring the TIR and FUV source luminosities, and then we relate this to the
extinction at other wavelengths as in \citet{2001PASP..113.1449C}:

\begin{equation}
A_{FUV}= 5.7\times \hbox{log}\left({1\over 1.68}{L_{TIR}\over L_{FUV}} +1\right)
\end{equation}

\begin{equation}
A_{NUV}= 0.72\times A_{FUV}
\end{equation}

\begin{equation}
A_V= 0.3\times A_{FUV}
\end{equation}

\begin{equation}
A_{H\alpha}= 0.25\times A_{FUV}.
\end{equation}

\begin{figure}
\includegraphics[width=\columnwidth]{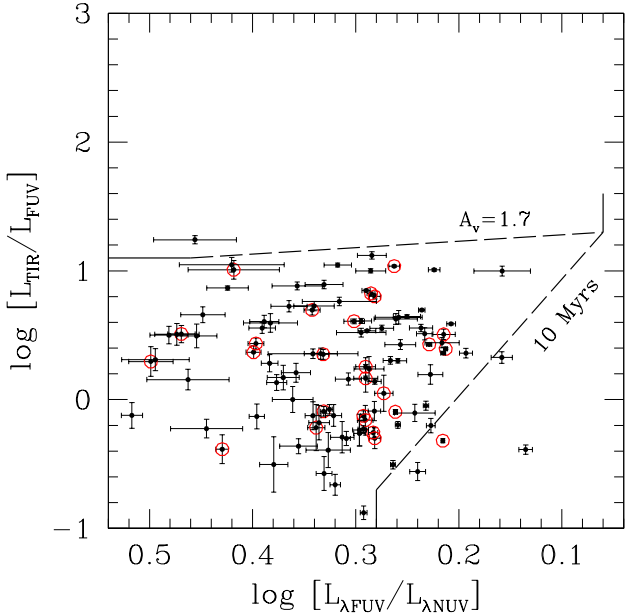}
\caption{Opacity expressed as infrared-to-FUV luminosity ratio versus
the observed UV color for all sources in the round sample.
Circles indicates sources in the isolated sample. All sources to the
left of the line marked 10~Myrs are younger than 10~Myrs according
to the instantaneous burst model. All
sources below the mark A$_V=1.7$ have visual extinction A$_V<1.7$~mag.}
\label{fig:av-col}
\end{figure}

We show in Figure \ref{fig:av-col} the ratio $L_{TIR}/L_{FUV}$ versus the GALEX
color $L_{\lambda}(FUV)/L_{\lambda}(NUV)$ for sources in the round sample.
By comparing this Figure
with Fig. 9 and 10 of \citet{2005ApJ...633..871C}, we see that our selected
sources have all  $\log \left(L_{TIR}/L_{FUV}\right)$ between $-1$ and 1.2,
corresponding to  $A_V < 1.7 $.
The derived value of $A_V$ increases on average towards bright sources.
The average UV color is on the order 0.3 mag.
All sources in the isolated sample and nearly all sources in the round sample
have UV colors $>0.2$ mag., i.e. they are younger than 10~Myrs
according to the instantaneous burst model for solar metallicity and for
a starburst dust distribution \citep{2000ApJ...533..682C}. Exact ages
will depend on the metallicity,  burst model and dust distribution.
An extinction curve like the one observed in the Small Magellanic Cloud,
as well as the simultaneous presence of newly formed massive stars
and an older UV-emitting population,
imply even younger ages for H$\alpha$ emitting sources in
our selected star-forming regions.

\begin{figure}
\includegraphics[width=\columnwidth]{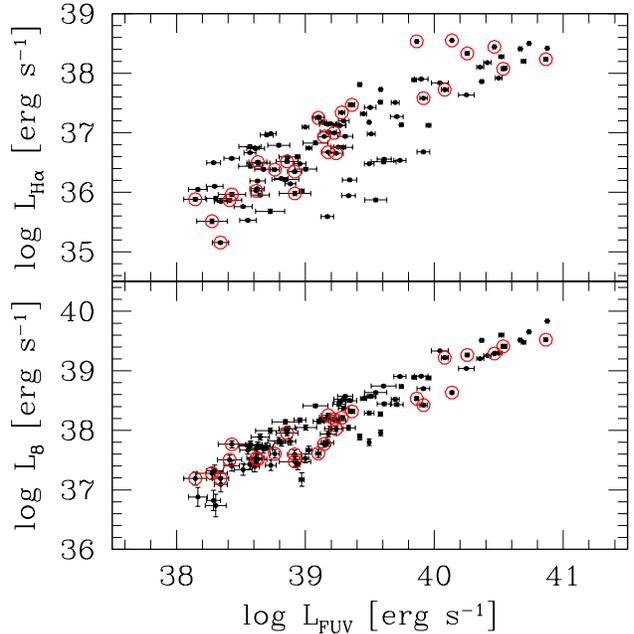}
\caption{H$\alpha$ and 8\mi\ luminosities are shown as functions of the
FUV luminosities for all sources in the round sample. Circles
indicate sources in the isolated sample. Luminosities are in units of
erg s$^{-1}$; L$_{H\alpha}$ and L$_8$ have been corrected for
extinction using the source TIR-to-FUV luminosity ratio.}
\label{fig:FUV-8-ha}
\end{figure}

We estimate the bolometric luminosity of our sources as:

\begin{equation}
L_{bol} = L_{TIR} + [\nu_{FUV} L_\nu(FUV) + \nu_{NUV} L_\nu(NUV)]
\label{eq:boluv}
\end{equation}

\noindent
where $\nu_{FUV}$ and $\nu_{NUV}$ are the effective frequencies of the
GALEX bands and the UV luminosities are uncorrected for extinction.
The bracketed quantity estimates the cluster UV luminosity, between 912
and 3000~\AA. In the absence of dust the UV luminosity is a good
approximation of the bolometric luminosity for
clusters with  ages between 1-10~Myr, independent of their mass
\citep[as from Starburst99][]{1999ApJS..123....3L}
\citep[see also][for similar methods used
to estimate L$_{bol}$]{2005ApJ...625...23B,2007ApJS..173..572T}.
We shall give another proof in the next Section
that indeed our selected clusters fall into this age category.
When dust absorb part of the UV radiation and re-irradiate it in
the IR, L$_{TIR}$ is needed in the right hand side of the above equation
to correctly estimate L$_{bol}$.

We apply extinction corrections to the FUV and H$\alpha$ luminosities
of sources in the round sample according to the model described earlier
in this Section.
The good correlation between extinction corrected UV luminosities
and 8\mi\ luminosities is evident in Figure~\ref{fig:FUV-8-ha}.
This is expected since PAHs are excited by the UV radiation of the
nearest star-forming region. This
nice correlation confirms the goodness of our extinction corrections.
\citet{2007ApJS..173..572T} have found that
PAH emission at 8\mi\ is suppressed  within strong star-forming regions.
The linear correlation between extinction corrected far-UV luminosities
and 8\mi\ luminosities, which does not break at high
luminosities, suggests that this is not happening in M33.

A correlation between FUV and H$\alpha$ luminosities is also present.
Given the different dependencies of these two luminosities on cluster
mass and age, the presence of some scatter in their relation is
not surprising. The next section will better address this issue.

%%%%%%%%%%%%%%%%%%%%%%%%%%%%%%%%%%%%%%%%%%%%%%%%%%%%%%%%%%%

\section{The cluster birthline: from bright to dim candles}

%%%%%%%%%%%%%%%%%%%%%%%%%%%%%%%%%%%%%%%%%%%%%%%%%%%%%%%%%%%

Because we resolve stars
and clusters in M33 at IR and optical wavelengths when the brightest member
is earlier than a B2 type star, the
expected relation between IR and H$\alpha$ emission is far from
linear even under the circumstance that most of the
bolometric luminosity of massive stars is absorbed by dust and re-emitted
in the IR. This is a result of the different dependencies of the
bolometric and H$\alpha$ luminosities on the stellar mass
\citep{1973AJ.....78..929P,1996ApJ...460..914V,2003ApJ...599.1333S}, 
considering the lack of the highest mass stars in some clusters.
In this Section we would like to investigate in detail the emission
by individual clusters at different wavelengths  to
judge the reliability of using infrared and H$\alpha$ emission as a 
quantitative tracer of recent star formation for a range of luminosities.

\subsection{Modelling cluster birthlines}

To compute the H$\alpha$ luminosity of a stellar cluster, one needs to
know the mass distribution at the upper end of the IMF. Even though
the stellar IMF is well known for intermediate stellar masses,
where the original exponent derived by \citet{1955ApJ...121..161S}
is now widely tested and used, uncertainties remain at the high-
and low-mass ends.

The IMF may be interpreted as a probability distribution
function that specifies the probability for a randomly chosen
star to lie within a certain mass range
\citep[e.g.][ for stochasticity in the IMF]{2006A&A...451..475C,2008A&A...484..711B}.
The absence of stars with very high masses suggests there is
a fundamental upper limit to the IMF 
\citep[$>120-150$~M$_\odot$ see][]{1998ApJ...493..180M,2000ApJ...539..342E,
2005Natur.434..192F,2005ApJ...620L..43O,2006MNRAS.365..590K,2008ApJ...675..163H}.
The most massive clusters sample the IMF out to this maximum possible stellar mass 
and no further, even though there are enough stars to do so if the IMF continued 
with the same slope.
Low and intermediate mass clusters do not generally have stars near the maximum 
possible mass, but their most massive stars tend to have a mass that increases 
with the cluster mass \citep{1982MNRAS.200..159L}. This trend leads to the question of 
whether the upper limit to the stellar mass in a particular cluster depends on the 
cluster mass because of some physically limiting process, or whether it follows only
from randomly sampling the IMF. In the first case, low mass clusters could not
make high mass stars. In the second case they could, as long as there is enough gas, 
and intermediate-mass clusters should occasionally be found with unusually massive stars 
-- ``outliers'' in the IMF. An important difference arises for the summed IMF from many 
clusters: it should be steeper than any cluster IMF in the
first case and the same as the average cluster IMF in the second case. The summed IMF 
is the basis for the present day mass function in whole galaxies and is therefore 
important to understand for population models and studies of galaxy evolution.
A recent debate in the literature suggests the question about maximum stellar mass
should be settled observationally
\citep{2004MNRAS.348..187W,2006MNRAS.365.1333W,2006ApJ...648..572E}.

In order to model
the distribution of stellar masses within a young cluster,
we follow two different approaches using the same IMF.
In one case we assume that each
cluster has a maximum stellar mass that is explicitly related to
the cluster mass. We call this the
{\it maximum mass case}. In the second case, we assume that stars of all masses
can form in clusters of all masses, provided there is enough gas, and that
the choice of a particular stellar mass is random, following the IMF. We call this
the {\it randomly sampled case}.
Once we specify how a cluster is built we need to establish
$L_{bol}^{cl}$ and $L_{H\alpha}^{cl}$, the
cluster bolometric and H$\alpha$ luminosity.
This is done by summing the contributions of all stars in the cluster
using stellar masses, Ly-continuum, and bolometric luminosities
tabulated
by \citet{1996ApJ...460..914V} for stellar masses 87.6$<$M$<$19.3~M$_\odot$ and
by  \citet{1973AJ.....78..929P} for 19.3$<$M$<$8~M$_\odot$. We extrapolate
luminosity values logarithmically if stellar masses
are outside these mass ranges and interpolate logarithmically for
masses that are intermediate to those shown in their tables.
We assume ZAMS (Zero Age Main Sequence) and solar metallicity and
divide the number of Ly-continuum photon
rates by $7.3\times10^{11}$~ph~erg$^{-1}$ to obtain H$\alpha$ luminosities.

In what follows we compute the theoretical {\it cluster birthline}
in the plane log($L_{bol}/L_{H\alpha}$)-log($L_{bol}$) using a stellar IMF with a 
Salpeter slope $\alpha=-2.35$ between 1 and 120~M$_\odot$. No stars with mass
lower than 1~M$_\odot$ or higher
than 120~M$_\odot$ will be considered. The IMF in stellar clusters
extends down to masses lower than 1~M$_\odot$ but these stars
give a negligible contribution to the cluster luminosity, which is
the relevant quantity for this Paper.

In the {\it maximum mass case}, given a number N of cluster members with
masses above 1~M$_\odot$, there will be a certain mass limit $M_N$ above
which the probability of finding one star is unity and this star will
be of mass $M_*$.
If $\xi(M)$ is the number of stars produced per unit mass,

\begin{equation} \label{eq:imf}
\xi(M)dM=C M^{\alpha} dM,
\end{equation}

\noindent
then the number of cluster members with masses above 1~M$_\odot$, and
the cluster mass and luminosity contributed by those stars, can be
written as

\begin{equation} \label{eq:cb1}
 N_{cl}^{>1}=\int_1^{120} C M^{\alpha} dM = 1+
{\int_1^{M_N} M^{\alpha} dM \over \int_{M_N}^{120} M^{\alpha} dM}
\label{eq:n}
\end{equation}

\begin{equation} \label{eq:cb2}
M_{cl}^{>1}= \int_1^{120} C M^{\alpha+1} dM = M_* +
{\int_1^{M_N} M^{\alpha+1} dM \over \int_{M_N}^{120} M^{\alpha} dM}
\end{equation}

\begin{equation} \label{eq:cb3}
L_{cl}^{>1}= \int_1^{120} C L(M) M^{\alpha} dL = L(M_*) +
{\int_1^{M_N} L(M) M^{\alpha} dL \over \int_{M_N}^{120} M^{\alpha} dM}.
\label{eq:l}
\end{equation}

\begin{figure}
\includegraphics[width=\columnwidth]{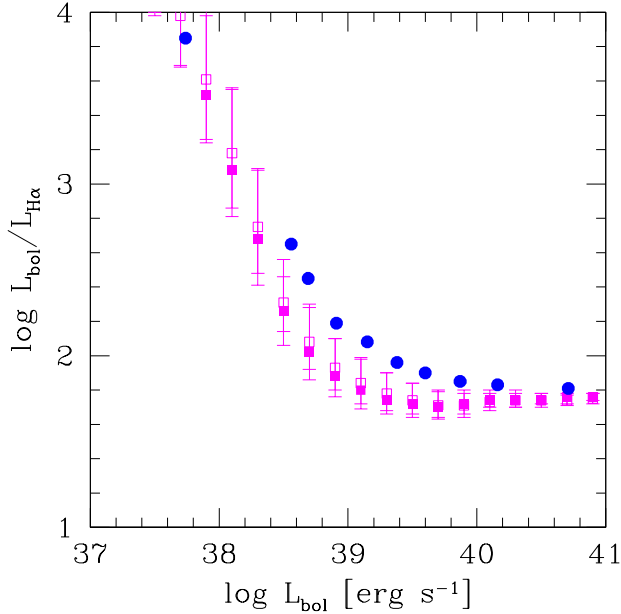}
\caption{Filled circles mark the expected values of the bolometric-to-H$\alpha$
luminosity ratio for the {\it maximum mass case}. Open and filled squares
are the mean and median values respectively of the same ratio for the {\it randomly
sampled case} using a cluster mass function index of $\delta=-2$. Errorbars point 
out the dispersions around the mean and median values.}
\label{fig:imf1}
\end{figure}

\noindent
We computed the normalization constant $C$ by assuming that the
number of stars between 120~M$_\odot$ and M$_{N}$ is unity and
we considered the IMF from M$_{N}$ to 1~M$_\odot$ as a continuous
distribution function with $\alpha=-2.35$
\citep[e.g.][ and references therein]{2006MNRAS.365.1333W}.
We computed the cluster masses
and luminosities in this way for a sequence of clusters with maximum stellar masses
corresponding to ZAMS stars of spectral types B2, B1, .... O4, and O3.  We set M$_N$ 
equal to the mass of these stars
and consider also that M$_*$,  the mass of the star born in the mass range
M$_N$--120~M$_\odot$, is equal to the lower limit of this range, M$_N$.
Statistically M$_{N}$ is the most likely
value for mass in the interval M$_N$--120~M$_\odot$, but the assumption
that for a given cluster mass
no stars are ever made with M$>$M$_N$ is stronger.
In Figure \ref{fig:imf1} we plot with filled circles the {\it cluster birthline} 
made with the maximum mass assumption.
That is, each filled circle corresponds to a spectral type in the above sequence,
with the corresponding bolometric-to-H$\alpha$ cluster luminosity ratio on the ordinate
and the bolometric luminosity on the abscissa for clusters that have a maximum stellar 
mass with that spectral type.

\begin{figure}
\includegraphics[width=\columnwidth]{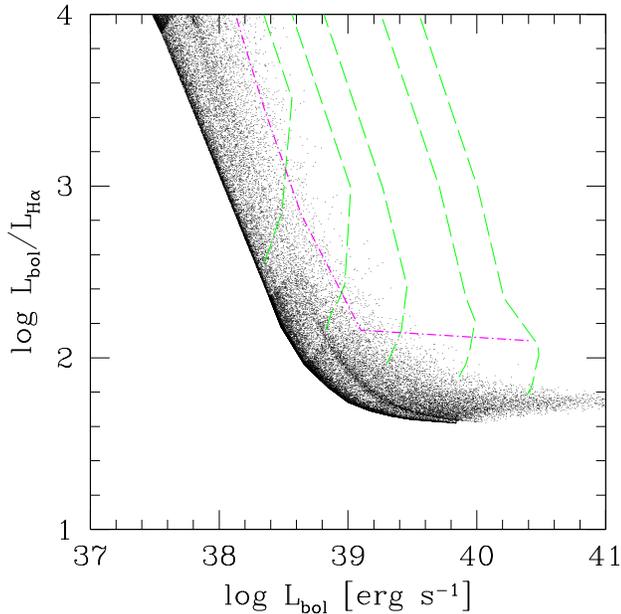}
\caption{The distribution of the ratio of the bolometric-to-H$\alpha$
luminosity versus the bolometric luminosity for newly born clusters simulated
in the {\it randomly sampled case}. Vertical dashed lines
mark evolutionary sequences, i.e., aging moves clusters upward from the simulated
birthline because H$\alpha$ luminosities fade away more rapidly than
bolometric luminosities. Clusters reach the top of the plot
(L$_{bol}$/L$_{H\alpha}\simeq4$)
in about 10~Myrs. The dash-dotted lines are indicative of when the most
massive stars in a cluster leave the main sequence.}
\label{fig:imf2}
\end{figure}

In the {\it randomly sampled case},
there is no explicit relationship between cluster mass and maximum stellar mass
as in equations (\ref{eq:n})-(\ref{eq:l}).
Even though M$_N$ is the most
likely mass for the cluster's most-massive member, the actual mass can be as large as
120~M$_\odot$, although the probability that such a massive star actually forms
decreases as the cluster mass decreases.
To model this case, we simulate 40000 clusters that
are distributed in number according to their mass between 20
and 10000~M$_\odot$ as

\begin{equation}
N(\hbox{ M}_{cl}) \hbox{ dM}_{cl} = \hbox{ D}_{cl} \hbox{ M}^\delta_{cl} \hbox{ dM}_{cl}
\end{equation}

\noindent
where $D_{cl}$ is a constant and $\delta$ is the spectral index of the ICMF, usually assumed
to be $\delta=-2$ \citep[e.g.][and references therein]{2007ChJAA...7..155D}.
We then populate each cluster with stars, randomly selected
from the stellar IMF described above, until the simulated cluster mass
gets above the selected cluster mass. We compute the bolometric and
H$\alpha$ luminosity of each cluster using the same ZAMS models as in the 
{\it maximum mass case}.
Figure \ref{fig:imf2} shows the resulting distribution of L$_{bol}$/L$_{H\alpha}$ 
versus L$_{bol}$ with one plotted point for each modelled cluster. 
The lower boundary of the distribution is where single massive stars
lie. Our simulation takes into account incomplete clusters, such as
single OB associations born in the field or single stars
scattered out of loose clusters. Notice the presence of a second edge
of high density points in the distribution.
This is from a change in slope of the stellar L$_{H\alpha}$-L$_{bol}$
relation, which produces a cusp in the diagram. In bins of equal size in
$\log$ L$_{bol}$ units, we  plot in Figure \ref{fig:imf1} the average and the median of
the distribution for the {\it randomly sampled case} and we plot their
1-$\sigma$ values (enclosing 16.5\% and 83.5$\%$ of the points in each bin).
Notice that the {\it randomly sampled case} predicts median values and
average values of log L$_{bol}$/L$_{H\alpha}$ that are below those of
the {\it maximum mass case}. This is a result of the occasional presence of
bright stars or outliers even in low-mass clusters.

If low mass stars form first, a cluster enters the
birthline from the top left and follows it down and to the right as 
L$_{bol}$ increases. If, instead, massive
stars are born at random times during cluster formation,
then the cluster can jump down on the simulated birthline and then move
towards slightly higher values of L$_{bol}$ and L$_{bol}$/L$_{H\alpha}$.
This happens until all of the gas available to fuel star formation is used 
or SF is quenched by feedback.
Cluster aging from the death of massive stars increases
the L$_{bol}$/L$_{H\alpha}$ ratio and then the cluster moves upward from the birthline.
This is because the death of massive stars makes the cluster H$\alpha$
luminosity fade away more rapidly than bolometric luminosity.
Using the public code {\it Starburst 99}, we evaluate how the clusters evolve off
the birthline. The vertical dashed lines in Figure \ref{fig:imf2} show the
aging effect. A cluster reaches the top of the Figure in about 10 Myrs.
Two fundamental properties of the predicted cluster
birthline are: $(a)$ the linear correlation between L$_{bol}$ and L$_{H\alpha}$,
which holds for very young massive clusters (the horizontal part of the birthline), 
breaks down below a critical value of
the cluster bolometric luminosity, depending on the stellar
IMF at the high mass-end; $(b)$
all stellar clusters should lie on or above the birthline shown
in Figure~\ref{fig:imf1} for the two cases examined.

\subsection{The M33 test case for the birthline}

\begin{figure}
\includegraphics[width=\columnwidth]{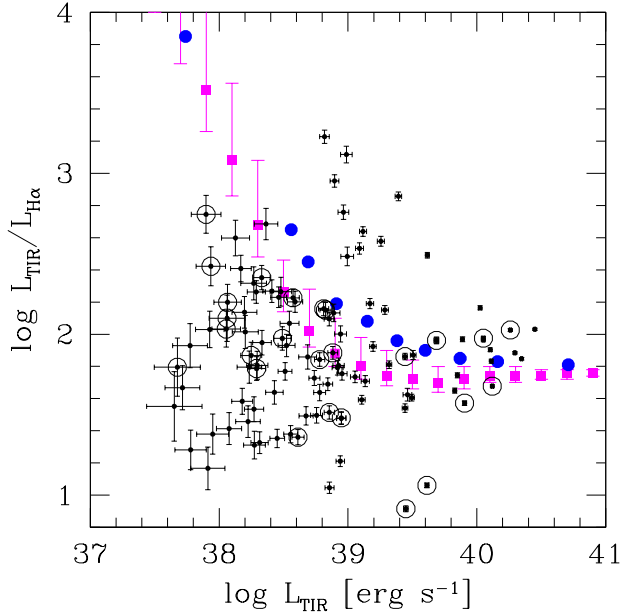}
\caption{The total infrared-to-H$\alpha$
luminosity ratio for data in the round sample and in the
cleanest isolated sample (symbols surrounded by circles). Under the
assumption that L$_{TIR}$=L$_{bol}$, we can compare the cluster data with
the predicted {\it cluster birthlines}
for the {\it randomly sampled case} (filled squares) and for the
{\it maximum mass case} (large filled circles). Individual extinction corrections
have been applied to H$\alpha$ luminosities (see Section 3).
The selected samples of young clusters is clearly not along or above
any birthline, as it should be. Hence the assumption L$_{TIR}$=L$_{bol}$
is incorrect.}
\label{fig:imf3}
\end{figure}

If most of the luminosity of  young clusters were absorbed by
dust and re-irradiated in the IR, then the L$_{TIR}\simeq$ L$_{bol}$
data points in the log(L$_{TIR}$/L$_{H\alpha}$) -- log(L$_{TIR}$) plane should all lie
above the birthline  sequence.  However, by looking at Figure \ref{fig:imf3}, 
which plots the observed values on the birthline plane,
we see a clear discrepancy with the birthline theoretical prediction. Many
clusters lie below the birthline, for both the maximum mass case and the
randomly sampled case. It has to be pointed out that
any heavier extinction correction, as well as corrections due to the loss
of ionizing photons leaking out
of the HII regions, will move the data points further down, toward lower
values of L$_{bol}$/L$_{H\alpha}$. Since any data point
below the birthline is unexplained we must conclude
that our assumption, namely that the TIR luminosity traces the bolometric
luminosity, is not correct for most of the clusters observed.  This discrepancy 
with the birthline cannot be the result of an unreliable 24\mi\ - TIR conversion 
because the dispersion in the  24\mi\ to TIR luminosity ratio is  only about 0.1 
in the log. Also recall that individual extinction
corrections have been applied to H$\alpha$ luminosities (see Section 3).

\begin{figure}
\includegraphics[width=\columnwidth]{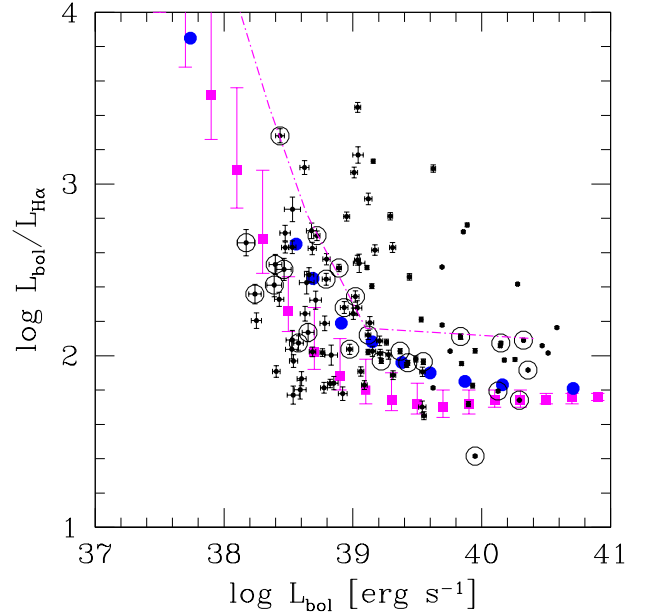}
\caption{Bolometric to H$\alpha$
luminosity ratio versus bolometric lumosity for data in the round sample 
and in the cleanest isolated sample (dots
surrounded by circles). Bolometric luminosities have been estimated 
according to Eq. \ref{eq:boluv} and individual extinction corrections
have been applied to H$\alpha$ luminosities (see Section 3). 
The predicted {\it cluster birthlines}
for the {\it randomly sampled case} (filled squares) and for the
{\it maximum mass case} (filled circles) are also shown.
The dash-dotted line indicates the cluster position when the most
massive stars for the randomly sampled case leave the main sequence.}
\label{fig:imf4}
\end{figure}

As outlined in the previous Section, we now complement IR with UV photometry
to derive the cluster bolometric luminosity.
In Figure \ref{fig:imf4} we plot the $L_{bol}$-to-$L_{H\alpha}$ ratio as a function
of $L_{bol}$ for the round sample marking (with circles) members of the isolated
sample. Clusters in
the round and isolated samples lie above the birthline predicted by the
{\it randomly sampled case} with a very few exceptions. Members of the isolated sample,
for which UV photometry is more accurate, follow closely the birthline and
this underlines their young age. Only one cluster in the isolated sample with
L$_{bol}\simeq 10^{40}$~erg~s$^{-1}$ seems to lie below the birthline,
having a lower value of L$_{bol}$/L$_{H\alpha}$.
The most likely explanation of the H$\alpha$ excess is that
this cluster has formed a star more massive
than 120~M$_\odot$, but additional analysis is needed before drawing any
definitive conclusion on this.
In general the agreement is good also for the round sample: almost all of the
clusters lie above the birthline for the randomly sampled case. This
implies that the round sample is made up of relatively young clusters too.
In Figure \ref{fig:imf4} we  add the main sequence
boundary (dash-dotted line). This
marks the value of  L$_{bol}$/L$_{H\alpha}$ for a cluster at the end of the main sequence
lifetime for single massive stars \citep[e.g. ][]{2005fost.book.....S}.
It is impressive how the isolated sample is bounded by the
birthline at the bottom and by the main sequence boundary at the top.
That means that effectively we have traced the sequence of young compact
clusters before supernovae disrupt them and blow away dust grains.

If the birth of stars in clusters is
regulated by the statistical character of the IMF and there is no
physical or absolute relation between the maximum stellar mass in a cluster and
the cluster mass, then the predicted
cluster birthline is in good agreement with the data.
This agreement is illustrated by
the relative number of sources below the birthline, where there should be none. Only
5$\%$ of the clusters are more than 3-$\sigma$ below the birthline in the
randomly-sampled case, whereas 25\% of the clusters are more than
3-$\sigma$ below the birthline in the maximum-mass case.
This result
favours the randomly sampled model for the
stellar population in clusters.

We infer that some of the clusters with the lowest 
L$_{bol}$/L$_{H\alpha}$ ratios for their L$_{bol}$ values
have outlier massive stars, i.e. massive stars without their usual proportion of 
low mass stars.  In these cases, a
star has formed with a mass that is larger than the average upper mass
limit for a cluster of that mass. It would be interesting to study these cases more
to see if there are other unusual characteristics, such as extreme mass segregation,
cloud disruption, or cloud temperatures.

\begin{figure}
\includegraphics[width=\columnwidth]{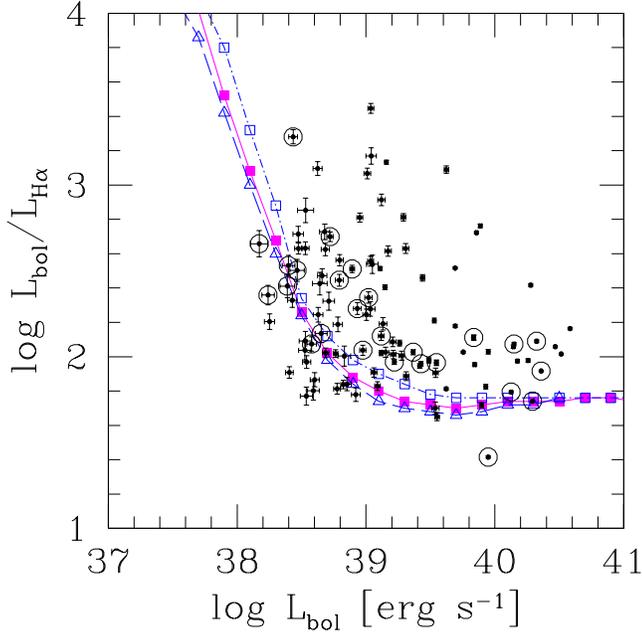}
\caption{Median values for the randomly sampled case obtained using
different indices for the cluster distribution function:
open squares, filled squares
and open triangles correspond to $\delta=-1,-2,-3$ respectively.
Filled dots show the data for the round sample. Data for the cleanest 
isolated sample are surrounded by circles.}
\label{fig:imf5}
\end{figure}

\begin{figure}
\includegraphics[width=\columnwidth]{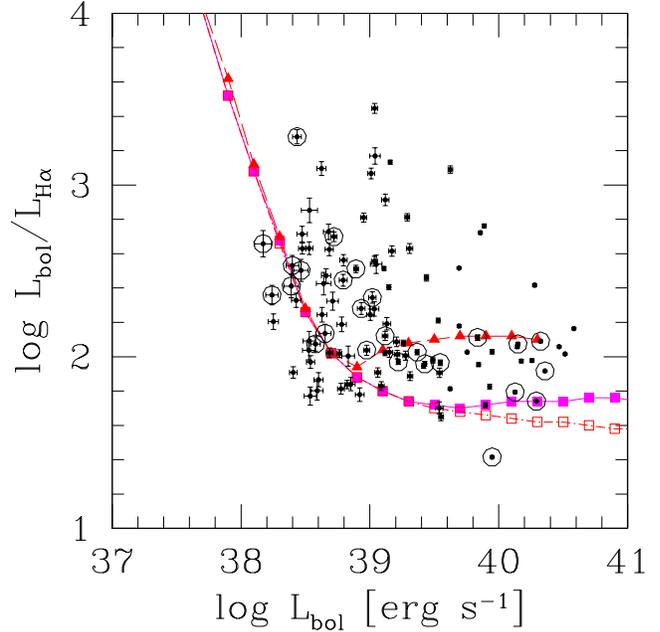}
\caption{Median values for the randomly sampled case obtained using
different index for the stellar IMF: filled triangles, filled squares
and open square correspond to $\alpha=-2.80,-2.35,-1.50$.
Filled dots show the data for the round sample. Data for the cleanest 
isolated sample are surrounded by circles.
}
\label{fig:imf6}
\end{figure}

The simulated L$_{bol}$/L$_{H\alpha}$  distribution is very marginally dependent
on the  value of $\delta$ for the cluster mass function. For $-1\le \delta \le -3$ 
the distribution of
the median value is within the 1-$\sigma$ error bar values of the more
widely used $\delta=-2$ distribution (see Figure \ref{fig:imf5})
\citep{2003AJ....126.1836H,2006ApJ...650L.111C,2008AJ....135..823D}.
Figure \ref{fig:imf6} shows
the effect of using an IMF index steeper or flatter than the classical
Salpeter value. There is a better agreement of some cluster data with a steeper
IMF at the high-mass end, but the number of clusters below the birthline increases.
So, we favor a Salpeter IMF with some cluster aging to bring them above
the birthline.  A diagram of L$_{bol}$ versus UV colors
shows that the most luminous sources are also the oldest in our isolated
sample. This accounts for the slight rise in plotted points above the birthline for 
large L$_{bol}$ in Figure \ref{fig:imf4}.  Additional data is
needed to better constrain the ages of those clusters.

%%%%%%%%%%%%%%%%%%%%%%%%%%%%%%%%%%%%%%%%%%%%%%%%%%%%%%%%%%%%%%%%%%%%%%%%%%%%%%%

\section{Summary and Discussion}

%%%%%%%%%%%%%%%%%%%%%%%%%%%%%%%%%%%%%%%%%%%%%%%%%%%%%%%%%%%%%%%%%%%%%%%%%%%%%%%

M33 is ideal for deriving the properties of young star-forming
clusters and their massive stellar populations. It contains star-forming complexes
with a wide range of luminosities, and it is close enough that even clusters
with one O- or B-type star can be localized.
We have examined in this paper young stellar clusters selected
in the 24\mi\ map of M33 that have H$\alpha$ counterparts.
The 8-to-24\mi\ luminosity ratio has a small dispersion and
shows a dependence on the galactocentric radius.
The total IR-to-H$\alpha$ luminosity ratio,
L$_{TIR}$/L$_{H\alpha}$,  shows a larger dispersion, especially towards
faint sources,
and no correlation with galactocentric distance or source IR luminosity.
The large scatter in the TIR-to-H$\alpha$ flux
arises mostly because of variations in the local dust abundance which adds to 
some aging effect.
Only a certain fraction of the bolometric luminosity of 24$\mu$m-selected sources
can be absorbed by grains and re-emitted locally at IR wavelengths. This fraction
might be small for low-luminosity sources that are not born along
the spiral arms of M33.

We have calculated the bolometric luminosity for 106 ``round'' stellar
clusters, L$_{bol}$, using UV and IR photometry. This has allowed us to
test the concept of the {\it cluster birthline} introduced in this paper.
The cluster birthline, defined in the parameter space
log(L$_{bol}$)--log(L$_{bol}$/L$_{H\alpha}$), is the line of birth of
young clusters,  a theoretical lower boundary for the ratio 
L$_{bol}$/L$_{H\alpha}$ for each L$_{bol}$.
We show that along the cluster birthline the relation between
L$_{bol}$ and L$_{H\alpha}$ is linear for high-L$_{bol}$ clusters but
is non-linear for clusters with L$_{bol}< 3\times 10^{39}$~erg~s$^{-1}$. 
Deviations from linearity are expected because of two effects: the IMF is not fully 
sampled in low mass clusters, and the
number of photons emitted by a star with energies above the hydrogen ionization
threshold has a stronger dependence on
stellar mass than does the stellar bolometric luminosity
\citep[e.g.][]{1973AJ.....78..929P,1996ApJ...460..914V}.
For high L$_{bol}$ clusters, the IMF is usually sampled out to the highest possible
stellar mass and then the ratio of high-to-low mass stars is constant, making the
L$_{H\alpha}$ to L$_{bol}$ ratio constant. For low L$_{bol}$ clusters, the IMF becomes
depleted of the highest mass stars; these stars have the lowest
ratio of L$_{bol}$/L$_{H\alpha}$ and their depletion raises the average ratio.
The observed flatness of  L$_{bol}$/L$_{H\alpha}$ for luminous sources
implies either that there is a maximum possible stellar mass equal to $\sim120\;M_\odot$,
as assumed in our models, or that the L$_{H\alpha}$-to-L$_{bol}$ ratio for individual stars
becomes constant at $M>120\;M_\odot$.

We modeled the cluster birthline by populating clusters of various
masses with stars having masses selected from an IMF. Several cluster mass
function slopes and IMF slopes were used for comparison. We modeled the
upper end of the IMF in two cases. The {\it maximum mass case} assumed that each
cluster produced stars with masses only up to a maximum value that equals the
average maximum stellar mass for a cluster of that mass. In this case, low mass clusters
can produce only low mass stars.  The {\it randomly sampled case} assumed that
each cluster can produce stars over the full range of the IMF, in which case a low
mass cluster can occasionally produce a high mass star (provided the cluster mass
exceeds the stellar mass, of course). Both cases have the same average maximum stellar
mass for clusters of each mass, but only the second case can produce outlier stars --
stars with masses significantly above the average maximum for that cluster, which means
above the upper end of the smooth and declining part of the IMF that is observed
at intermediate stellar mass.  Comparing the two cases, the clusters birthline in the 
maximum mass case has a higher value of L$_{bol}$/L$_{H\alpha}$ over the range of L$_{bol}$
where O- and B-type stars are starting to populate the upper end of the IMF. This range extends
from L$_{bol}$ slightly less than $10^{38}$~erg~s$^{-1}$ to L$_{bol}$ slightly larger
than $10^{40}$~erg~s$^{-1}$. At smaller L$_{bol}$, stars producing H$\alpha$ emission
do not form in the maximum mass case and are highly unlikely in the randomly sampled case.
At larger L$_{bol}$, the IMF is fully sampled in both cases.

From the round sample we selected 26 sources which appear isolated and
compact in the UV maps 
i.e. the annular region around them, used for background 
subtraction, is not contaminated by UV sources.  
All but one of the stellar clusters in this sample, called the isolated sample 
are compatible with 
cluster birthline predicted in the randomly sampled case for the Salpeter IMF 
with maximum stellar mass limit of 120~M$_\odot$.
For the isolated sample in fact 25 out of 26 clusters lie on or above this birthline.
Considering the whole round sample we find similar fractions:
100 over 106 clusters are compatible with the same birthline.
In contrast, 25\% of the clusters in the round sample lie below the theoretical 
birthline in the maximum mass case, and hence are incompatible with this cluster 
population model. M33 has therefore provided a positive test to the cluster
birthline concept introduced in this paper. Stellar clusters are born
along the birthline and aging moves the clusters above it.
The observations also suggest that
stars randomly sample the IMF and that clusters occasionally have outlier massive stars.
Clusters in the isolated sample lie below the upper main sequence boundary
($t\le 3-4$~Myr), which is consistent with their compactness and young age. Below this 
boundary, the most massive stars that a cluster formed are still present (assuming 
that all stars form at about the same time), so L$_{bol}$/L$_{H\alpha}$ remains low, and
supernovae have not yet disrupted the cluster or blown
away the dust. No cluster in the isolated sample lies
above the upper main sequence boundary, but some clusters in the more numerous
round sample do. This difference might indicate that some members of the round sample have
lost their most massive stars.
However, in the opacity--UV-color diagram, the two samples occupy the same
area so their ages are about the same.
An alternative explanation for the high L$_{bol}$/L$_{H\alpha}$ ratio is then
leakage of ionizing photons from HII regions.
Clusters in the round sample are in fact not as compact and isolated as the
clusters in the isolated sample, so the non-isolated clusters could be more prone to leakage.
Future studies that resolve individual
stars should help to answer this age-versus-leakage question.
If massive stars are observed directly, and in proportional to the IMF, then leakage must be
the cause of the higher L$_{bol}$/L$_{H\alpha}$ ratio.

Luminosities at 8 and 24\mi\ correlate with gas metal abundances;
the dispersion is large because there are
few sources with an accurate metal abundance determination. 
While we did not find a significative sample of 24\mi\ sources embedded in
GMCs which are not detected in H$\alpha$, we find 34 sources
associated with GMCs which have an H$\alpha$ counterpart.
The 24\mi\ luminosities of these sources 
correlate with associated GMC masses, even though with a large scatter.

From studies in the solar neighborhood we know that a stellar cluster in its
youngest stage is likely to be deeply embedded in molecular clouds. Later,
mechanical and radiative effects of
the most massive stars disrupt the cloud, the extinction decreases and the cluster
becomes visible in the ultraviolet. The time interval in which massive
stars are detectable through IR, H$\alpha$ and UV emission
depends not only on stellar lifetimes but also on the opacity of the
parent cloud and on its
time evolution. The characteristics of molecular
clouds seem linked to the large-scale morphology of the galaxy.
Earlier studies of a radio-selected sample of thermal sources in M33
have provided 11 young, optically-visible stellar clusters but no embedded
star cluster \citep{2006ApJS..162..329B}. A small sample of compact infrared
selected sources with no H$\alpha$ counterpart has been observed in the
CO J=1-0 and J=2-1 lines (Corbelli et al. 2009, in preparation). The
weakness or absence of CO lines suggests that these sources are not
embedded proto-stellar clusters in the process of formation but a more
evolved population.
In M33 the amount of extinction seems generally lower than in the Milky Way.
The absence of large molecular complexes and a steeper mass spectrum
implies that even in an early SF phase many
stellar complexes may not be highly obscured. Later,
winds and SN explosions remove efficiently the dusty envelope
and the cluster fades away in the IR before less massive stars
get off the main sequence.

The concept of a {\it cluster birthline} together with the high resolution of
future telescopes seems a promising way to analyze
star formation in external galaxies.

\begin{acknowledgements}
We would like to thank R. Walterbos for providing us
the H$\alpha$ image of M\,33, R. Bandiera, L. Hunt,
P. Lenzuni, and F. Palla for stimulating discussion
on the Cluster Birthline and the referee, R. de Grijs, for his criticism 
to an earlier version of the paper.
The work of S.~V. is supported by a INAF--Osservatorio Astrofisico
di Arcetri fellowship.
This research has made use of Spitzer Space Telescope data and of
GALEX mission data. We acknowledge Spitzer Space Telescope center operated
by the Jet Propulsion Laboratory, California Institute of Technology,
under contract with the National Aeronautics and Space Administration,
and NASA's support for the construction, operation, and science analysis
of the GALEX mission, developed in cooperation with the Center National
d'Etudes Spatiales of France.
\end{acknowledgements}

\appendix

%%%%%%%%%%%%%%%%%%%%%%%%%%%%%%%%%%%%%%%%%%%%%%%%%%%%%%%%%%%%%%%%%%%%%%%%%%%%%%%

\section{The metallicity and GMC samples}

%%%%%%%%%%%%%%%%%%%%%%%%%%%%%%%%%%%%%%%%%%%%%%%%%%%%%%%%%%%%%%%%%%%%%%%%%%%%%%%

We now investigate if properties of star-forming sites in M33
are related to the host gas metallicity or to the parent molecular cloud mass.
We first consider a sample of HII regions observed by \citet{2007A&A...470..865M}
for which metallicities have been determined via optical spectroscopy.
Extinction can be estimated via the Balmer decrement, even though this is
only an upper limit to the overall extinction through a whole HII region.
Unfortunately the sample of HII regions with known metallicity for which
radio continuum or Paschen lines are available for determining the extinction
deeply into the star-forming region is limited to 5 bright sources and therefore
the sample is not statistically meaningful.
Our metallicity sample is made of 31 HII regions with known O/H abundance,
Balmer decrement,
and accurate IR photometry at 8 and 24~$\mu$m. H$\alpha$ emission has
been corrected for extinction using the Balmer decrement and the relation
$A_{V} = 1.2 \times A_{\rm H\alpha}$. The extinction (which gives values
$A_V< 1.7$ in the whole sample) is comparable to what has
been derived with more accurate estimates. There is no trend of extinction with
metallicity or with gas column density that is associated with the source site.
Low column density gas hosts preferentially low metallicity HII regions.

\begin{figure}
\includegraphics[width=\columnwidth]{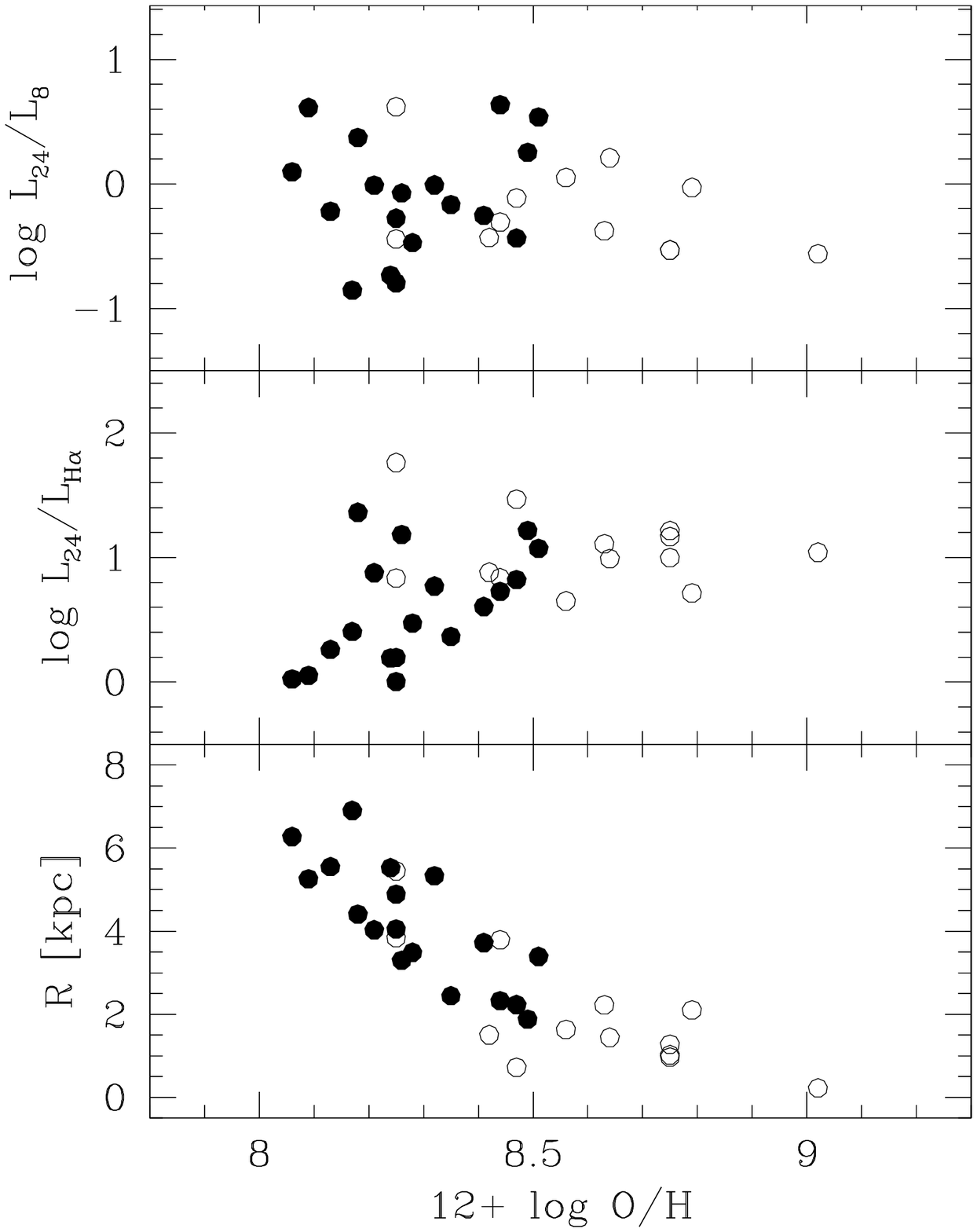}
\caption{Metallicity as a function of galactocentric radius for sources
with IR counterparts. Filled dots represent sources  where the electron
temperature diagnostic lines have been detected. The metallicity measured
in terms of O/H from a compilation by \citet{2007A&A...470..843M,2007A&A...470..865M}
is also plotted as a function of the ratios of the 24\mi\ to the H$\alpha$
luminosities, and the 24\mi\ to the 8\mi\
luminosities. Error bars in log(O/H) are given in
\citet{2007A&A...470..843M,2007A&A...470..865M} and are on the order 0.1.
Error bars in the luminosity ratios are smaller than the symbol sizes.
}
\label{app1}
\end{figure}

\begin{figure}
\includegraphics[width=\columnwidth]{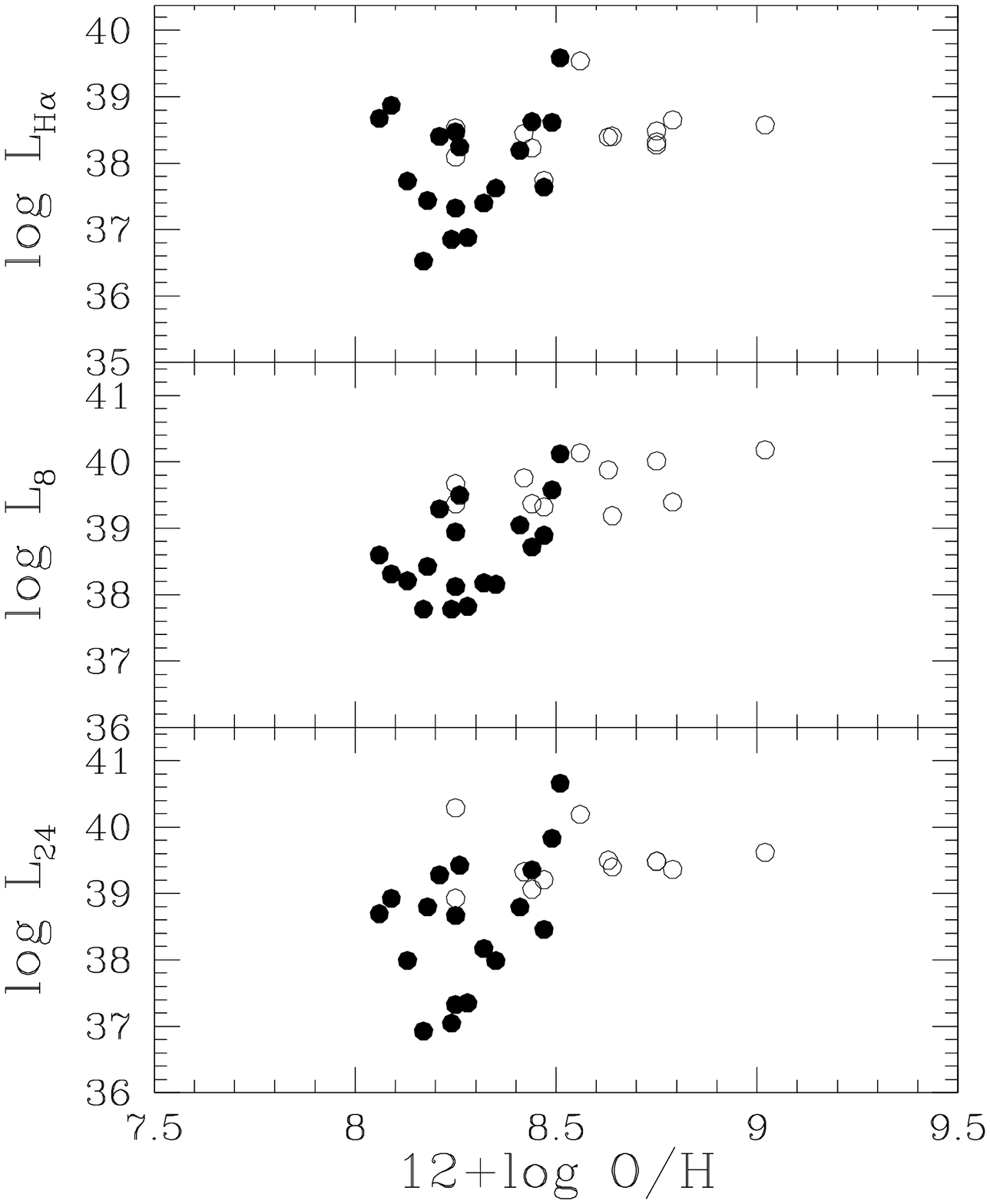}
\caption{The metallicity measured
in terms of O/H from a compilation of \citet{2007A&A...470..843M,2007A&A...470..865M}
is plotted as a function of the 24\mi, 8\mi, and H$\alpha$ luminosities for sources
with IR counterparts. Filled dots represent sources where the electron
temperature diagnostic lines have been detected.  Error bars in log(O/H) are given in
\citet{2007A&A...470..843M,2007A&A...470..865M} and are on the order 0.1.
Error bars in the luminosity ratios are smaller than the symbol sizes. }
\label{app2}
\end{figure}

The sample of sources with IR counterparts shows a shallow radial metallicity 
gradient (Fig. \ref{app1}).
If we restrict our sample to sources where the electron
temperature diagnostic lines have been detected, and hence the
electron temperature can be measured (filled dots in Figures \ref{app1}
and \ref{app2}), then the gradient is compatible with that derived by 
\citet{2007A&A...470..865M}
using the same selection criteria. The $L_{24}/L_{8}$ and
$L_{24}/L_{H\alpha}$ ratios do not show any
clear metallicity dependence because sources with high metallicity (say
O/H above $-3.5$)
have high luminosities at optical and IR wavelengths (Fig. \ref{app2}). 
The dependence is however very marginal if we consider only sources with electron
temperature determinations, whose metallicities are more certain.
The scatter around the mean values increases
as the metallicity  decreases, but this might be an effect of
large galactocentric radii, where most of the low metallicity
regions are found.

IR luminosities show a dependence on metallicity.
Bright sources at 8 and 24\mi\ are born where metallicity is high while
low metallicity
gas host a wider range of source luminosities. However uncertainties in the
metallicities of bright sources that have no detections of the electron
temperature diagnostic lines are large and do not allow us to
draw firm conclusions. 

\begin{figure}
\includegraphics[width=\columnwidth]{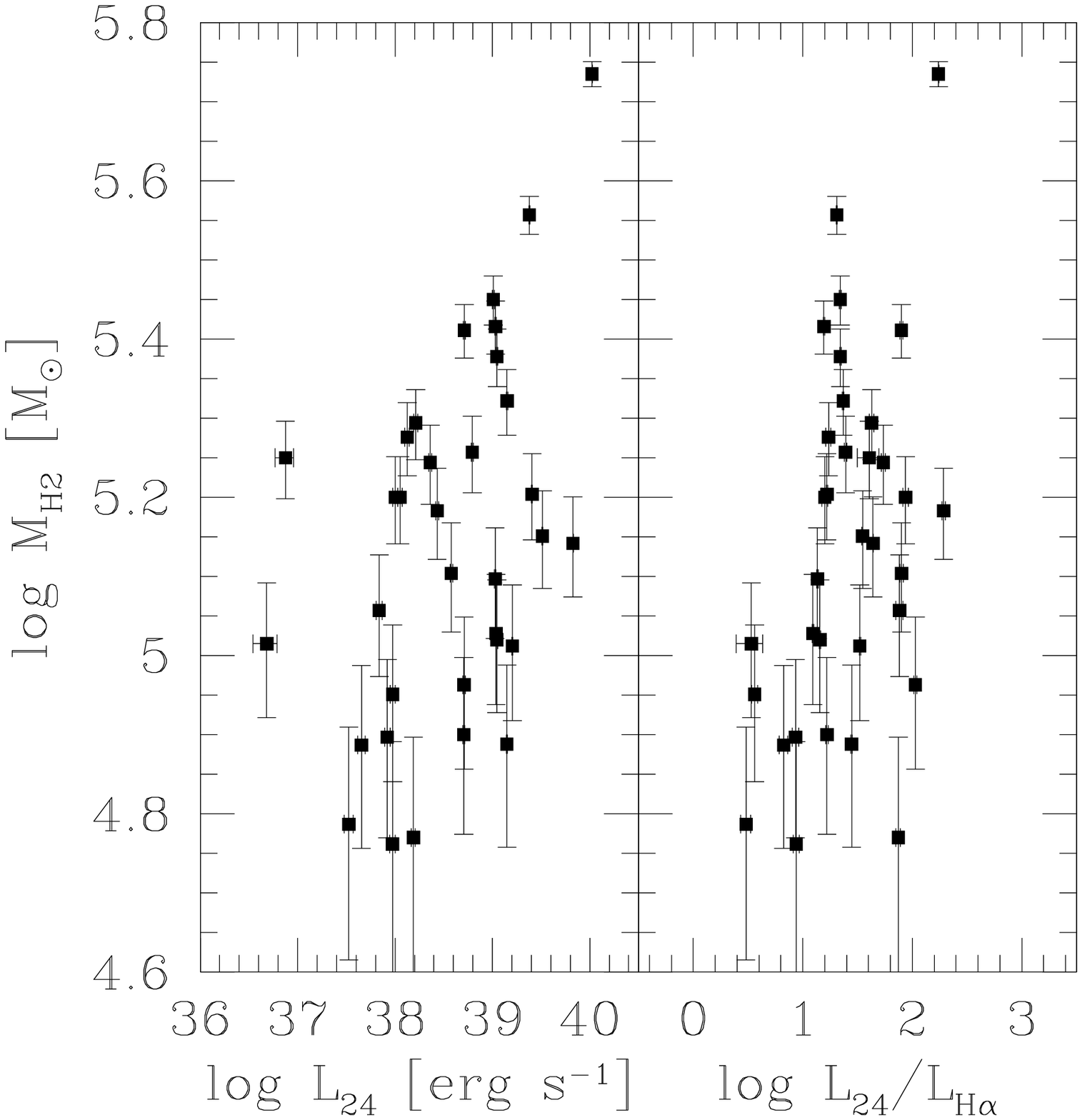}
\caption{The molecular cloud masses associated with the IR selected
sources are shown versus the 24\mi\ luminosities and and the ratios 
between this luminosity and the H$\alpha$ luminosity.
}
\label{app3}
\end{figure}

We also analyze the BIMA giant molecular cloud dataset
\citep[$13''$ spatial resolution;][]{2003ApJS..149..343E} to see
if any sources in the main sample are associated with GMCs and
in these cases if the source luminosities are related to the parent GMC masses.
Figure \ref{app3} shows  the GMC masses as functions of the $L_{24}$ and  
$L_{24}/L_{H\alpha}$ values for sources within 6.5 arcseconds of a GMC.
This limiting radius was chosen to
match the BIMA spatial resolution of the \citet{2003ApJS..149..343E} GMCs survey
that was used to identify the clouds.
There is a dependence of the IR source luminosity on the GMC mass, despite the
large scatter. The slope of the correlation is 0.17$\pm 0.05$ with Pearson linear
correlation coefficient of 0.53. The ratio $L_{24}/L_{H\alpha}$ for the GMC sample
varies by about 1.5 order of magnitudes for sources associated with small clouds,
and by even less for sources associated with the most massive clouds.
Sources in the GMC sample are then confined to a smaller range of $L_{24}/L_{H\alpha}$
than that reported in Figure \ref{app3} for the full sample. The ratio
$L_{24}/L_{H\alpha}$ shows no clear dependence  on the GMC mass.

%\begin{thebibliography}{}
\bibliography{astroph}
%\end{thebibliography}

\end{document}